\documentclass[11pt]{article}
\usepackage[top=1in, bottom=1in, left=1in, right=1in]{geometry}
\usepackage{graphics,graphicx}
\usepackage{caption}
\usepackage{amssymb,amsmath,amstext}
\usepackage{color}
\usepackage{times}
\usepackage[numbers,sort&compress]{natbib}
\usepackage{booktabs}
\usepackage{setspace}

\usepackage{flushend}
\usepackage{multirow}

\newcommand{\lyxaddress}[1]{
\par {\raggedright #1
\vspace{1.4em}
\noindent\par}
}

\begin{document}
\title{Solving Cram\'er-Rao Lower Bound in Single PMU Channel for Forced Oscillations in  Power Systems}
\author{Zikai Xu $^{\text{}1}$ and John W. Pierre $^{\text{}1}$}
\date{}
\maketitle
\lyxaddress{1. Department of Electrical and Computer Engineering, University of Wyoming}

\newpage
\begin{abstract}
Forced oscillations threaten the reliability of wide-area power systems, and different approaches to estimate forced oscillation have been explored over the past several years.  Though these efforts provide powerful tools to estimate a forced oscillation’s amplitude, frequency, and phase, a benchmark for estimation accuracy has not been available.  This paper provides initial results in such a benchmark called the Cram\'er-Rao  Lower Bound which is a lower bound on the variance in those estimates. A direct approach to derive  Cram\'er-Rao Lower Bound of forced oscillation in power systems is given. In the end, the MinniWECC model is applied to assess the impact of different factors on the bound.
\end{abstract}

\newpage
\section{Introduction}
    In power systems, a primary concern is the reliable delivery of electrical power; however,  power systems are sophisticated networks, consisting of generators and loads, spread over a wide geographical area. Power oscillations have evolved into a major problem threatening the reliability of wide area power systems. In the past decade, detection and suppression of power oscillations have been more of a focus in the power engineering community. As a result of the increasing number of synchronized Phasor Measurement Units (PMUs), collecting synchrophasor measurements from wide-area power systems, PMUs provide numerous data to study oscillations for improving power systems reliability. 

 Forced oscillations (FOs) are the response of the system associated with a periodic undesired external input, and thus the oscillation frequency is determined by the frequency of the periodic excitation \cite{ft2016}. FOs have been observed in the US Eastern Interconnection , the western North American Power system and the Nordic power system \cite{llj2011}. Compared with natural oscillations, which are continuously excited by random variation of loads or network suddenly switching \cite{wpt2003,kpt2016,fta2017}, FOs may have large amplitude and are frequently an indication of a problem in the power system needing to be addressed.  If the operating condition is poorly damped and the frequency of the forced oscillation is near the modal frequency, FOs can lead to blackouts \cite{hyp2016}. Numerous situations can cause FOs in power system; for instance, steam-turbine regulator malfunction \cite{wll2010} or incautious power system stabilizer design \cite{mf1990}. Thus, given PMU measurements, it is important to develop methods to detect and estimate FOs. Bonneville Power Administration (BPA) has significant experience in detecting forced oscillations and examples are provided in \cite{kla2016}.

FOs typically exist as sinusoids, characterized by amplitude, phase and frequency\cite{kp2019}. Methods have been developed to estimate amplitude, frequency and also phase of FOs as well as the location of origin of FOs\cite{fp2015,kp2019,ap2019,cm2012,mw2017}. Accurate estimation of the frequency of the forced oscillation has been shown to be critically important to a new class of mode meters.  Mode-meters estimate the electromechanical modes from ambient data.  When forced oscillations are present in the data, traditional mode meters \cite{ptd1997,tdz2008,wpt2003} may give biased estimates of the frequency and damping ratio if the frequency of the forced oscillation is close to the frequency of an electromechanical mode.  This limitation was addressed with the development of a new class of mode meters \cite{fpm2016,sfp2018} designed to be robust to the presence of forced oscillations.  These new mode-meters utilize an estimate of the forced oscillation frequencies.  In \cite{l2020,fae2021}, these algorithms were seen to benefit from highly accurate estimates of forced oscillation frequencies.  Thus it is important to understand how well those frequencies can be estimated. Through all these previous works, the true values of  parameters provide a benchmark; however, a benchmark for parameter standard deviation has not been provided. Alternatively, among all these algorithms, a standard deviation of estimated parameters are compared to each other, and estimated parameters, generally, are close to the true value; but a lower bound on the standard deviation has not been studied.


This paper focuses on finding a lower bound on the standard deviation (STD) of the estimated frequency, amplitude, and phase of forced oscillations.  The lower bound is called the Cram\'er-Rao Lower Bound (CRLB), which gives insight into the limits as to how well forced oscillation parameters can be estimated. This is important for insight into the estimate of the amplitude of the forced oscillation but even more important for insight into the estimate of the frequency of the forced oscillations since these new generation of mode meters depend on accurate estimates of frequency.  In this paper, a direct method is derived to solve CRLB which clearly indicate the key variables influencing the estimation of the amplitude, frequency and phase.  These variables include the record length, the power spectral density of the ambient noise, and the amplitude of the forced oscillation.The importance of the power spectral density has also been observed in the detection of forced oscillations \cite{kp2018, fp2015}. The scope of the paper focuses on results for a single measurement channel and does not address the detection of forced oscillations.


\emph{Notation:} The set of integer number is denoted by $\mathbb{Z}$. The set of real number is denoted by $\mathbb{R}$. A constant or a parameter is indicated
by non-bold letters($A$,$\theta$).  Matrices or vectors are
denoted by bold letters ($\boldsymbol{C}$,$\boldsymbol{\theta}$). Further, transpose of a vector $\boldsymbol{\theta}$ is
given by $\boldsymbol{\theta}^{\top}$ and the N-dimensional identity matrix is denoted by $\boldsymbol{I}_{N\times N}$. In addition, subscript $w$ denotes a parameter in white noise case, $A_{w}$.

The paper is organized as follows. Section \ref{sect2} introduces
 modeling of a FO in ambient data. A direct method to derive CRLB of FO's amplitude, phase and frequency is given in the Section \ref{sect3}. Section \ref{sect4} demonstrates an example to apply the method and discusses the results. Section \ref{sect5} concludes the paper.

\section{Mathematical Model for Forced Oscillation in Ambient Data}\label{sect2}

In a power system, the distributed effect of random load variations act as a continuous excitement of the power system dynamics  \cite{wpt2013}.  Such randomness can be modeled as approximately white noise over the frequency band of interest.  This can be viewed as the power system being excited by the random load variations $w[n]$ producing colored noise $v[n]$ as shown in Fig. 1.  The colored power system noise is referred to as ambient noise and therefore contains information about the system dynamics.
e
 The time domain structure of $v[n]$ can be represented by the power spectral density (PSD) function $S_{v}(f)$, providing that mean-squared-value of a signal is spread across frequency, in the frequency domain. For example if one takes the frequency deviation from a phasor measurement unit (PMU) as the signal of interest and subtracts off the mean value, then the PSD describes how the variance of the frequency deviation is spread across frequency.  The total area under the power spectral density equals the variance in the frequency deviation.  The area under the PSD between two frequencies is the contribution to the overall variance in the frequency deviation coming from that frequency band. It is well known that the PSD of ambient power system data tends to have peaks near the frequencies of the electromechanical modes.
 
  
\begin{figure}[tbph]
\centering 
    \includegraphics[width=0.4\textwidth]{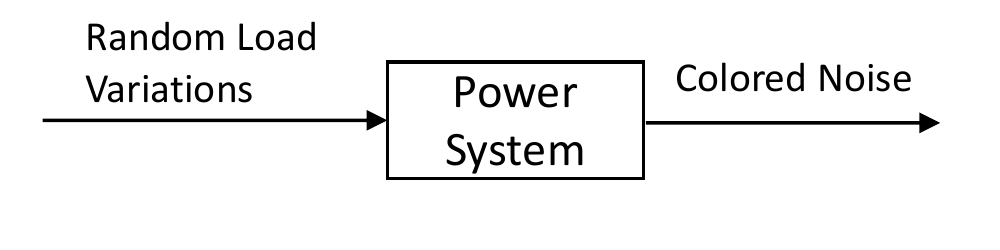}
\caption{Power system with white noise input and colored noise output}\label{systemplot1}
\end{figure}

\begin{figure}[tbph]
\centering 
\includegraphics[width=0.45\textwidth]{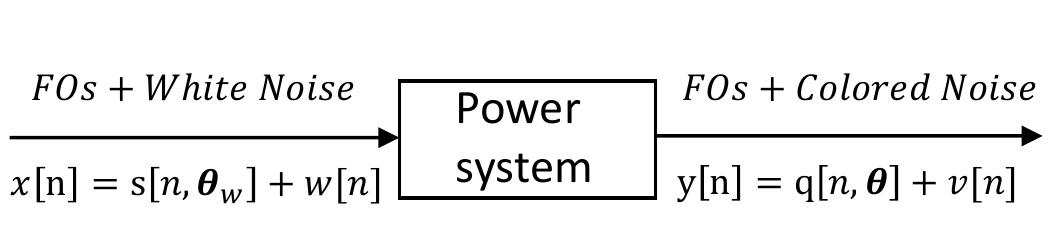}
    \caption{Model sketch of the power system}\label{systemplot}
\end{figure}

When forced oscillations are present in the PMU measurements , 
the sampled input signal becomes the sum of white noise data and the forced oscillation:
\begin{equation}\small
    x[n] = s[n] + w[n], n=0,1,\dots,N-1,
\end{equation}
where $x[n]$ is input data; $w[n]$ is white noise and $N$ is the length of estimation window or measurement window. The term $s[n]$ represents any FOs in the signal. Because of  appearance of FOs in PMU measurement as sinusoids \cite{luj2018}, it is appropriate to give a FO term as sinu
\begin{equation} \small
    s[n,\boldsymbol{\theta}_{w}] = A_{w}\cos{(2\pi f_{0}n+\phi_{w})},
\end{equation}
where $A_{w} \in \mathbb{R}$ is the amplitude of the oscillation. $f_{0} \in [-1,1]$ is the normalized frequency (i.e. $f_{0}=F_{0}/F_{s}$) where $F_{0}$, with unit $Hz$ (cycles per second), is the frequency of the forced oscillation, and $F_{s}$ is sampling frequency of PMU with unit of samples per second.  $\phi_{w}  \in [-\pi, \pi]$ is the phase in radians. Concisely, these parameters can be written as vector form  $\boldsymbol{\theta}_{w} = [{A_{w}} \:\: {f_{0}}\:\: {\phi_{w}}]^{\top}$.
Considering that FO and white noise are independent, the steady state output $y[n]$ can be found as 
\begin{equation}
    y[n] =q[n,\boldsymbol{\theta}] + v[n], n=0,1,\dots,N-1
\end{equation}
where $v[n]$ is colored noise or called as ambient noise of the power system. $q[n,\boldsymbol{\theta}]$ is FO; precisely, the responded FO of the power system, which can be written as 
\begin{align}
        q[n,\boldsymbol{\theta}]  =   {A}\cos{(2\pi {f_{0}}n+{\phi})},
\end{align}
where $\boldsymbol{\theta} = [{A} \quad {f_{0}}\:\: {\phi}]^{\top}$. ${A}$, ${f_{0}}$ and ${\phi}$ are the
 amplitude, frequency and phase, respectively, of FO in the power system to be estimated. The system sketch is given in Fig.\ref{systemplot}.

In this section, the model for FO in ambient noise is introduced. Our previous works  \cite{fpm2016,kp2018,fp2015,sfp2018,kp2020,luj2018,fp2015con} present a variety of approaches to estimate parameters in FO; however, the lowest variance of estimating parameters under study among them. In the next section, Cram\'er-Rao Lower Bound is explored to overcome this shortcoming.

 \section{Cram\'er-Rao Lower Bound} \label{sect3}
 Voluminous PMUs are allocated in the power system to collect synchrophasor measurements data of voltage and current.  Based on PMU measurements, one can have different algorithms to estimate the amplitude, phase, and frequency of a FO \cite{fpm2016}; thus, a benchmark should be proposed to evaluate different estimation algorithms. The CRLB is used as a classic benchmark for assessing estimator performance. CRLB is the lower bound of the variance of any unbiased estimator; in other words, CRLB is a bound on the smallest variance by any unbiased estimator. 
 
\subsection{Solving Fisher information matrix  }
To solve for the CRLB, the Fisher information matrix needs to found.  In this paper, as amplitude, frequency and phase are parameters to estimate, $\boldsymbol\theta=[A, f_{0}, \phi ]^{\top}$. The Fisher information matrix 
$\boldsymbol I(\boldsymbol \theta)$ can be defined as

\begin{equation}
    \boldsymbol{I}(\boldsymbol{\theta})=\begin{bmatrix}
           [\boldsymbol{I}(\boldsymbol{\theta})]_{11} & [\boldsymbol{I}(\boldsymbol{\theta})]_{12}& [\boldsymbol{I}(\boldsymbol{\theta})]_{13}\\
            [\boldsymbol{I}(\boldsymbol{\theta})]_{21} & [\boldsymbol{I}(\boldsymbol{\theta})]_{22}& [\boldsymbol{I}(\boldsymbol{\theta})]_{23}\\
             [\boldsymbol{I}(\boldsymbol{\theta})]_{31} & [\boldsymbol{I}(\boldsymbol{\theta})]_{32}& [\boldsymbol{I}(\boldsymbol{\theta})]_{33}
    \end{bmatrix}, \label{fishmatirx}
\end{equation}
and each entry in the matrix $\boldsymbol{I}(\boldsymbol{\theta})$ can be solved as 
\begin{equation} \small
    [\boldsymbol{I}(\boldsymbol{\theta})]_{ij}= -E\left[\frac{\partial^2 ln p[\boldsymbol{y},\boldsymbol{\theta}]}{\partial\theta_{i}\partial\theta_{j}} \right],\label{fisher}
\end{equation}
where $i,j\in\{1,2,3\}$, and $E$ stands for expectation value \cite{xu2020noise,xu2018exact,xu2019analysis,xu2021initial,xu2022cramer1,xu2022cramer} .The signal $y[n]$ has been defined in (4); since $v[n]$ is colored Gaussian noise, the probability density function of output $y[n]$ can be written as
\begin{equation}  \small
 p(\boldsymbol{y};\boldsymbol{\theta}) = \frac{1}{(2\pi)^{\frac{N}{2}}det^{\frac{1}{2}}[\boldsymbol{C}]}\exp{ \left\{ \frac{1}{2}\left(\boldsymbol{y}-q(\boldsymbol{\theta})\right)^{\top}\boldsymbol{C}^{-1}(\boldsymbol{y}-q(\boldsymbol{\theta}))\right\} }, \label{likelihood}
\end{equation}
where $\boldsymbol{C}\in \mathbb{R}_{N \times N}$ is the covariance matrix of the colored noise $v[n]$, which is 
\begin{equation}\small
    \boldsymbol{C}= E \left\{ \begin{bmatrix} 
    v[0]v[0] &  v[0]v[1] &\dots & v[0]v[N-1]\\
     v[1]v[0] &  v[1]v[1] & \dots&  v[1]v[N-1]\\
     \vdots   &  \vdots  & \ddots   & \vdots \\
    \scriptstyle   v[N-1]v[0] & \scriptstyle v[\small{N}-1]v[1]& \dots& \scriptstyle  v[\small{N}-1]v[\small{N}-1]
    \end{bmatrix} \right\}\label{Cmatrix},
\end{equation}
and $det[\cdot]$ implies determinant of the matrix $\boldsymbol{C}$.

Substitute \eqref{likelihood} into \eqref{fisher}, one can simplify the fisher information matrix as 
\begin{equation} \small
    [\boldsymbol{I}(\boldsymbol{\theta})]_{ij}= \left[\frac{\partial q(\boldsymbol{\theta}) }{\partial \theta_{i}} \right]^{\top}\boldsymbol{C}^{-1}\left[\frac{\partial q(\boldsymbol{\theta}) }{\partial \theta_{j}} \right],\label{fishersimply}
\end{equation}

where 

\begin{equation}\small
    \frac{\partial q(\boldsymbol{\theta}) }{\partial \theta_{1}}=
    \begin{bmatrix} 
  \cos{{\phi}}\\
    \cos{(2\pi {f_{0}}+{\phi})}\\
  \cos{(4\pi {f_{0}}+{\phi})}\\
    \vdots 
    \\
    \cos{\left(2(N-1)\pi{f_{0}}+{\phi}\right)}
    \end{bmatrix},  \label{theta1}
\end{equation}

\begin{equation} \small
    \frac{\partial q(\boldsymbol{\theta}) }{\partial \theta_{3}}=
    \begin{bmatrix} 
          -{A} \sin{({\phi})}\\
    -{A} \sin{(2\pi {f_{0}}+{\phi})}\\
    -{A}\sin{(4\pi {f_{0}}+{\phi})}\\
    \vdots 
    \\
    -{A}\sin{\left(2(N-1)\pi{f_{0}}+{\phi}\right)}
    \end{bmatrix}, \label{theta3}
\end{equation}
\begin{equation} \small
    \frac{\partial q(\boldsymbol{\theta}) }{\partial \theta_{2}}=
    \begin{bmatrix} 
          0\\
    -{A}2\pi \sin{(2\pi {f_{0}}+{\phi})}\\
    -{A}4\pi\sin{(4\pi {f_{0}}+{\phi})}\\
    \vdots 
    \\
    -{A}2(N-1)\pi\sin{\left(2(N-1)\pi {f_{0}}+{\phi}\right)}
    \end{bmatrix}. \label{theta2}
\end{equation}

\subsection{Solving CRLB}
By applying CRLB theorem \cite{k93}, true value of estimation parameter is used; CRLB are the diagonal elements of the inverse of the Fisher information matrix  
\begin{equation}\small
    Var(\hat{\theta}_{i}) \geq \left[\boldsymbol{I}^{-1}(\boldsymbol{\theta}) \right]_{ii},\label{crlb}
\end{equation}
where $i =1,2,3$. $\boldsymbol I^{-1}(\boldsymbol{\theta})\in \mathbb{R}_{3 \times 3}$ is inverse of Fisher information matrix given in \eqref{fishmatirx}. $[\cdot]_{ii}$ stands for the element at the $i$th column and $i$th row of the  matrix. $Var(\hat{\theta}_{i})$ is to  place a bound on the variance of $i$th parameter in $\boldsymbol{\theta}$.


To replace \eqref{theta1}, \eqref{theta3}, \eqref{theta2} and \eqref{Cmatrix} into \eqref{fishersimply} and finally solve CRLB by applying \eqref{crlb}. This method can give a numerical value of CRLB of estimation amplitude, frequency and phase.

\begin{figure*}[t]
	\vspace{2mm}
\centering 
\includegraphics[width=\textwidth]{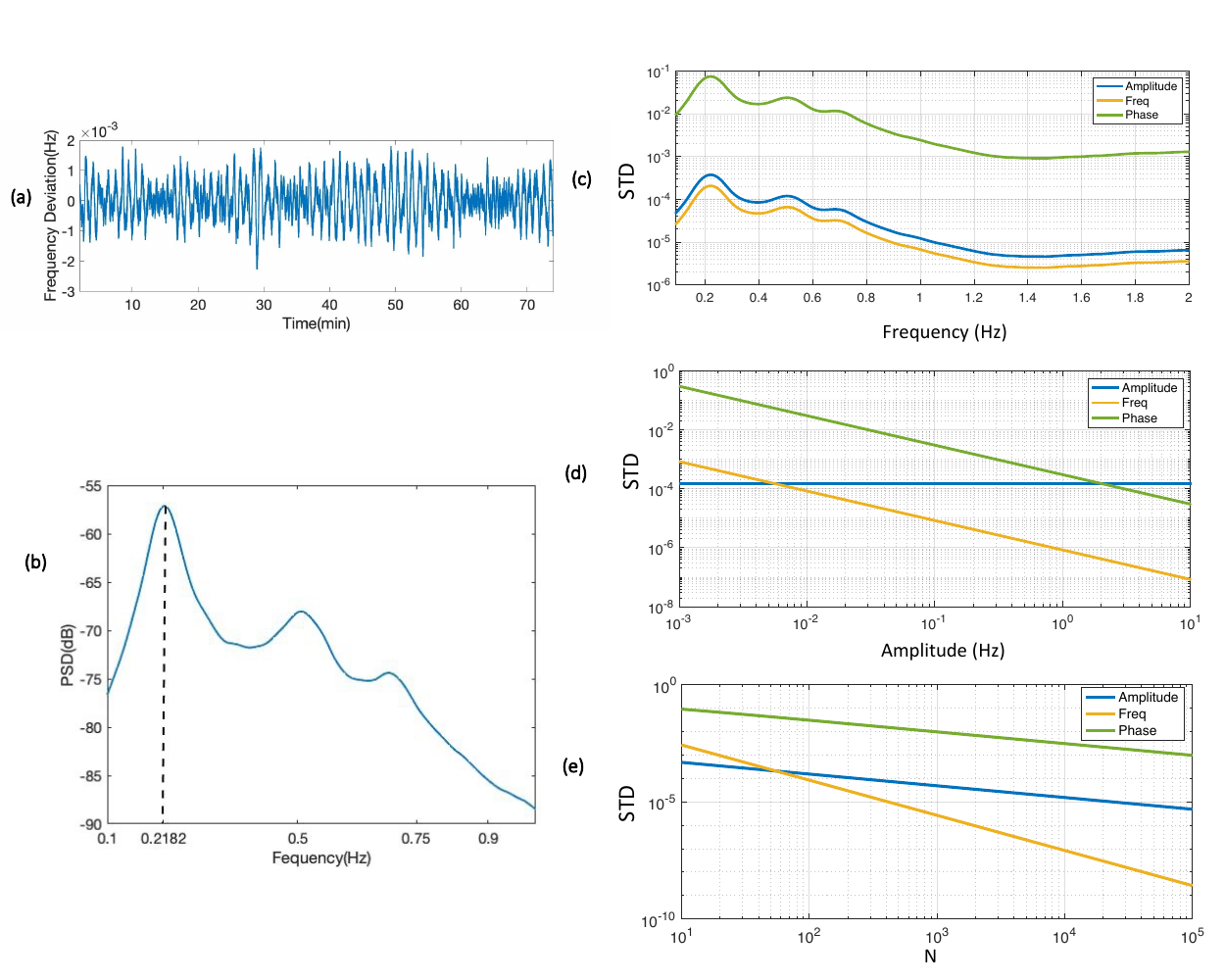}
    \caption{}\label{systemplotmain}
\end{figure*}

\section{Results and Discussions}\label{sect4}
In this section, MinniWECC model is applied to generate ambient power data. The MinniWECC model is a simplified model of the Western Electricity Coordinating Council (WECC) system where it has reduced order dynamics and is geographically consistent with the WECC system. In order to simplify the system, single generator is equivalent to generation for many areas, and MinniWECC only include transmission lines with significant length. Generally, the MinniWECC has 34 generators, 54 generator and load transformers, 115 lines and high-voltage transformers, 19 load buses, and 2 DC lines.
Details of MinniWECC specifications can be found in \cite{TU2008}. This nonlinear model is linearized about an operating point to generate data.

An example of generated data is given at Fig.3(a), where relative frequency is  calculated from bus 119 and bus 72 ( Bus 119 is up in Alberta Canada while bus 72 is in Colorado). The corresponding PSD estimate is shown in Fig.3(b). One can obviously observe that a peak in the PSD is in the the major inter-area mode of the model. Additionally, two low-energy modes near 0.6 and 0.7 Hz are visible in the PSD.

The method described in Section III is used to calculate the CRLB for various FO amplitudes, frequencies, and data record lengths. Fig. 3(c) gives the STD bound for amplitude, frequency, and phase as FO’s frequency increases. Here, record length of data is $100$, and sampling rate is $5$ samples per second; namely, $20$ seconds of observation data is tested. Since phase itself has not too much affect on the STD bound for amplitude, frequency, and phase; the phase value just randomly take $\pi/6$ radians.  Amplitude takes value of 0.01 $Hz$. We notice that CRLB is affected by system modes: when FO's frequency is close to system modes' frequencies, large variance of amplitude, frequency and phase estimation are expected.

Fig3.(d) plots STD bound for amplitude, frequency and phase as FO's amplitude varying.  Record length of data N is $100$. Phase value take $\pi/6$ radians. FO frequency is $0.3 Hz$. STD bound of amplitude itself keep same level at different amplitude value. STD bound of frequency and phase have similar decreasing rate when amplitude is increasing. Thus, generally speaking, large amplitude of FOs can improve estimation accuracy. 

Finally, in Fig3.(e)  the STD bound of amplitude, frequency and phase are decreasing  with increasing N number. Phase value take $\pi/6$ radians. FO frequency is $0.3 Hz$. Amplitude takes value of 0.01 $Hz$. Since more data points are available, estimation accuracy are increased; however, frequency estimation improve rapidly compared with the other two parameters.

\section{Conclusion}\label{sect5}

In this paper, a method to find the CRLB on the standard deviation for estimated FO amplitude, frequency and phase has been presented. Instead of focusing on estimating parameters in forced oscillation, the lower bound performance of estimation is explored. Cram\'er-Rao Lower Bound, known as a  lower bound of the variance of any unbiased estimator, is introduced. A method is given to solve Cram\'er-Rao Lower Bound: the approach directly gives the numerical value of bound. This method has its benefits and shortness: the approach is easy to apply, and does not need to have prior information of the power system; however, the numerical error can cause incorrect results. 
Finally, the MinniWECC model example is given to better understanding the CRLB on estimated parameters, and gives the reader insight to explore how different parameters affect estimation accuracy. In the future work,  closed form expressions for the CRLB will be determined to make them easier to compute and to more directly see the relationship between these standard deviation bounds and the parameters of the oscillation and the ambient noise.


\section*{Acknowledgment}

This work was supported by Sandia National Laboratory and funded by the U.S. DOE, Office of Electricity, Advanced Grid Modeling (AGM) Program. The authors are grateful to Dan Trudnowski of Montana Tech for use of the miniWECC model. The authors also appreciate the valuable discussions with Dave Schoenwald and Ryan Elliott of Sandia National Laboratory

\bibliography{RefMaster}
\bibliographystyle{unsrt.bst}      
	
\end{document}